\documentclass[prb,twocolumn,showpacs,preprintnumbers,amsmath,amssymb]{revtex4-1}
\usepackage{graphicx}
\usepackage{epstopdf}

\begin{document} 
\title{Effective medium theory for superconducting layers: A systematic 
analysis including space correlation effects} 
\author{S. Caprara$^1$, M. Grilli$^1$, L. Benfatto$^{1,2}$, and C. Castellani$^1$} 
\affiliation{$^1$Dipartimento di Fisica Universit\`a di Roma ``La Sapienza'', 
piazzale Aldo Moro 5, I-00185 Roma, Italy\\
$^2$Consiglio Nazionale delle Ricerche, Istituto dei Sistemi Complessi, 
via dei Taurini, I-00185 Roma, Italy}   
\begin{abstract} 
We investigate the effects of mesoscopic inhomogeneities on the metal-superconductor 
transition occurring in several two-dimensional electron systems. Specifically, as a 
model of systems with mesoscopic inhomogeneities, we consider a random-resistor 
network, which we solve both with an exact numerical approach and by the effective 
medium theory. We find that the width of the transition in these two-dimensional 
superconductors is mainly ruled by disorder rather than by fluctuations. We also 
find that ``tail'' features in resistivity curves of interfaces between LaAlO$_3$ 
or LaTiO$_3$ and SrTiO$_3$ can arise from a bimodal distribution of mesoscopic 
local $T_c$'s and/or substantial space correlations between the 
mesoscopic domains.
\end{abstract}   
\date{\today} 
\maketitle 
   
\section{Introduction}   
The recent discovery of superconductivity in transition metal oxide 
interfaces\cite{reyren,triscone,espci} and of pseudogap effects and 
disordered-induced inhomogeneity in thin conventional superconducting 
films\cite{sacepe1,sacepe2,sacepe3,mondal} has triggered a renewed interest 
in transport and superconductivity in two-dimensional electronic 
systems.\cite{review_feigelman} Owing to their low dimensionality and to 
unavoidable defects in the growing processes, disorder plays an important 
role in the physics of these systems. First of all, impurities lead to 
localization effects, which can degrade the superconducting critical 
temperature\cite{finkelstein} and can give rise to localized preformed 
Cooper pairs.\cite{leema,feigelman,dubi,randeria} As it has become evident 
in a number of recent numerical works,\cite{dubi,randeria} the system 
manifests a spontaneous tendency towards inhomogeneities even in the presence 
of a homogenous disorder distribution. In particular disorder can also give rise 
to patchy electronic systems which are inhomogeneous on a {\em mesoscopic} scale, 
with superconducting regions embedded in, or connected by, normal/insulating 
regions.\cite{ioffe,feigelman10,review_feigelman}

\begin{figure}   
\includegraphics[angle=0,scale=0.3]{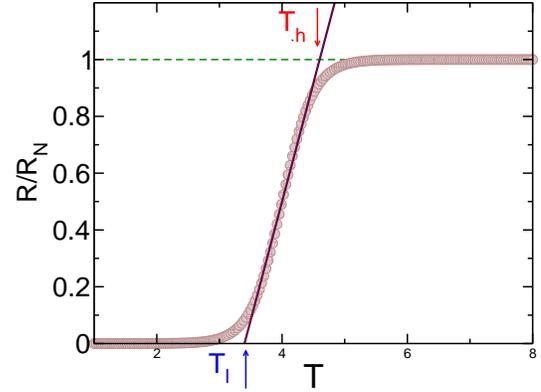}   
\caption{Sketch of the typical resistance curve for a superconducting
  interface (circles). The width of the transition can be estimated as
  $\Delta T_c/T_c\simeq (T_h-T_l)/T_l$. Notice the persistance of a
  pronounced tail also below $T_l$. By closer inspection of the data
 in Ref.\ \onlinecite{triscone}, one can see that for negative applied
  voltages $\lesssim -80$ V such a tail saturates to a low but finite
  resistance value.} 
\label{figdeltaT}   
\end{figure}

Understanding all such issues is a difficult task, particularly close to the 
superconductor-insulator transition, and it would obviously involve together 
concepts like percolation, establishment of coherence between superconducting 
``islands'', role of disorder in the Berezinski-Kosterlitz-Thouless transition, 
and so on. In this context it would be useful to disentangle the various aspects 
by understanding whether and which properties of these systems could be explained 
just in terms of mesoscopic inhomogeneities.  Therefore we find it important and 
timely to investigate systematically the effects of large scale inhomogeneities on 
the superconducting transition in two dimension, irrespectively of their microscopic 
origin. The analysis of these effects seems particularly compelling in metal-oxide 
interfaces, upon inspecting their sheet resistance $R_\square(T)$ data around 
the critical temperature. While homogeneously disordered films of conventional 
superconductors present rather sharp transitions with high slopes of 
$R_\square(T)$ around $T_c$,\cite{sacepe1,sacepe2,sacepe3,mondal} these 
superconducting heterostructures commonly display broad 
transitions\cite{triscone,espci} even for relatively small values of the 
normal-state resistance. The typical shape of $R_\square(T)$ near the transition 
is sketched schematically in Fig.\ \ref{figdeltaT}. In general, the downturn of 
$R_\square(T)$ towards zero is characterized by a linear regime with a relatively 
small slope. By defining the two scales $T_l$ and $T_h$ as in 
Fig.\ \ref{figdeltaT}, one can estimate approximately the width of the transition 
as $\Delta T/T_c\approx (T_h-T_l)/T_l$. The typical values for $\Delta T/T_c$ 
for superconducting interfaces are listed in Table I, along with the corresponding 
values in disordered films of conventional superconductors, as TiN,\cite{sacepe1} 
and NbN.\cite{mondal} 
Inspecting Table I one can easily see that the typical width of the superconducting 
transition in metal-oxide interfaces is substantially larger than the width of 
standard disordered films, that attain values comparable to those reported for 
the interfaces only at extremely large disorder concentration, where also the 
critical temperature is almost driven to zero. The observed temperature dependence 
of the sheet resistance in superconducting interfaces shows instead two 
characteristic features not reported in disordered films: (i) the transition width 
becomes large for relatively small variations of the critical temperature (compare 
the case $V=-40$ V and $V=-60$ V in Table I); (ii) an extended tail is observed 
below $T_l$ in a sample with an applied bias $V\lesssim -80 V$, with the 
resistance saturating to a low but still {\em finite} value.\cite{triscone} These 
two features call for a specific analysis to see whether in heterostructures 
these large transition widths and persisting tails arise from mesoscopic 
inhomogeneities.    
\begin{table}[t] 
\begin{center} 
\begin{tabular}{ccccc} 
\hline \hline 
 Ref. & Sample &  $T_h$ (K) &  $T_l$ (K) & $\Delta T/T_c$ \\ 
\hline 
[\onlinecite{triscone}]& V=180 V& 0.347 & 0.263 & 0.32 \\ 
 & V=100 V & 0.338 & 0.260 & 0.3 \\ 
 & V=0 V& 0.306 & 0.2 & 0.53 \\ 
 & V=-40 V & 0.26 & 0.138 & 0.88 \\ 
 & V=-60 V& 0.235 & 0.100 & 1.35 \\ 
\hline 
[\onlinecite{sacepe1}] &TiN1 & 1.71 & 1.34 & 0.28 \\ 
&TiN2 & 1.34 & 1 & 0.34 \\ 
&TiN3 & 0.83 & 0.47 & 0.76 \\ 
\hline  
[\onlinecite{mondal}]  & $k_F\ell =3.27$ & 8.27 & 8.50 & 0.03 \\ 
& $k_F\ell =1.58$ & 2.36 & 4 & 0.69 \\ 
\hline \hline 
\end{tabular} 
\end{center} 
\caption{Estimate of the superconducting transition width in
  superconducting interfaces (Ref.\ \onlinecite{triscone}) and disordered films of
  conventional superconductors (Refs.\ \onlinecite{sacepe1} for TiN
  and \onlinecite{mondal} for NbN).  The labels used to identify the samples
  are the same used in the corresponding publications. In the case of Ref.\
\onlinecite{triscone} $V$ is the bias potential applied to the SrTiO$_3$ substrate
to change the charge density at the interface.} 
\label{t-table} 
\end{table} 

It is worth noting that such broad transitions can hardly be ascribed to ordinary 
superconducting fluctuations. Indeed, considering the Aslamazov-Larkin\cite{AL} 
contribution of superconducting fluctuations to the paraconductivity [see 
Eq.\ (\ref{asllark}) below], one obtains that $\Delta T/T_c= R_N/R_c$, where $R_N$ 
is the normal-state sheet resistance (i.e., approximately the value right 
above the beginning of the downturn of $R_\square (T)$) and 
$R_c=16\hbar /e^2=65.6$ k$\Omega$. Since $R_N$ values for the interfaces are of 
order of 1 k$\Omega$ or even smaller, ordinary superconducting fluctuations would 
give $\Delta T/T_c\simeq 0.02$, i.e., at least one order of magnitude smaller than 
the one reported in Table I. This result has to be contrasted, for example, to 
other systems like high-temperature superconducting oxydes,\cite{vidal} where the 
conclusion was reached instead that fluctuation effects are enough to account for 
the width and rounding effects around $T_c$, while mesoscopic inhomogeneities play 
a minor role. 
  
The disordered mesoscopic model we have in mind consists of large metallic regions 
with randomly distributed critical temperatures that we map into a network of 
random resistors. Upon decreasing $T$ more and more regions (links) become 
superconducting until a percolation threshold is reached and the superconducting 
transition occurs. The model can  be solved by means of the so-called effective 
medium theory (EMT),\cite{effectivemedium,kirkpatrick} that is often 
used as a valuable guide to investigate the effects of mesoscopic 
inhomogeneities.  However, to establish the degree of reliability of EMT in the 
case of different disorder realizations, we will compare the EMT results with 
exact numerical solutions on finite clusters. In particular, owing to the 
``mean-field-like'' character of EMT, it is important to establish to what extent 
this approach can reliably be used in cases where space correlations are sizable.    

We notice that in our model the superconducting regions are assumed to be large 
enough to have a fully established local coherence and to make negligible the 
charging effects. As a consequence, neighboring superconducting resistances 
immediately establish a mutual phase coherence as soon as both have become 
superconducting. This clearly distinguishes our framework from the case of 
granular superconductors, where the grains have usually nanoscopic sizes of the 
order of the coherence length of the pure system. Moreover, the large size of a 
mesoscopic domain also allows to consider paraconductivity effects (both 
{\it \'a la} Aslamazov-Larkin\cite{AL} and {\it \'a la} Halperin-Nelson\cite{HN}) 
to occur within each domain.   

Our paper is structured as follows. In Sec. II we summarize some results of the 
EMT in the case of mixed normal-superconducting states. In Sec. III we describe 
our numerical approach, where we find an exact solution to a network of random 
resistors with different local resistances and/or critical temperatures. Sec. IV 
is devoted to the analysis of different effects ($T_c$ distribution, inclusion of
correlation between local $T_c$ and local resistance, gaussian or 
Berezinski-Kosterlitz-Thouless paraconductivity fluctuations) in the absence of 
space correlations. This last limitation is overcome in Sect. V, where different  
realizations of space correlations are considered. Our concluding remarks are 
reported in Sec. VI. 

\section{Effective medium theory}   
The effective medium resistivity $\rho_{em}$ of a random resistor network, in which 
the  value $\rho_i$ of the resistivity of a resistor on a bond of the network 
occurs with a frequency $w_i$, obeys the equation\cite{effectivemedium,kirkpatrick} 
\begin{equation}
\mathcal F(\rho_{em})\equiv\sum_i w_i\frac{\rho_{em}-\rho_i}{\rho_{em}+\alpha\rho_i}=0,
\label{effmed}
\end{equation} 
where the sum is carried over all the possible values of $\rho_i$, $\sum_i w_i=1$, 
and the parameter $\alpha$ is related to the connectivity of the network. For a 
cubic network in $D$ spatial dimensions, $\alpha=D-1$. Eq.\ (\ref{effmed}) can be 
recast in the self-consistent form 
\begin{equation}
\rho_{em}=\sum_i\frac{w_i\rho_i}{\rho_{em}+\alpha\rho_i}\left[\sum_i\frac{w_i}{\rho_{em}+\alpha\rho_i} 
\right]^{-1}\equiv\Phi(\rho_{em}),
\label{effmed1}
\end{equation} 
with $\rho_P\le \Phi(\rho_{em})\le\rho_S$, $\rho_P\equiv[\sum_i w_i/\rho_i]^{-1}$, 
$\rho_S\equiv\sum_i{w_i\rho_i}$, and $\Phi^\prime(\rho_{em})\ge 0$. It is therefore 
evident that the function $\Phi(\rho_{em})$ interpolates between the parallel 
$\rho_P$ and the series $\rho_S$ of the resistors. Since in Eq.\ (\ref{effmed}) 
$-1/\alpha\le\mathcal F(\rho_{em})\le 1$, and $\mathcal F^\prime(\rho_{em})>0$, 
Eq.\ (\ref{effmed}) has a unique solution, which can be efficiently found, e.g., 
by bisection, within the interval $[\rho_{min},\rho_{MAX}]$, where $\rho_{min}$ and 
$\rho_{MAX}$ are, respectively, the smallest and largest possible values of the 
resistivity. Indeed, from Eq.\ (\ref{effmed1}), it is evident that 
$\rho_{min}\le\rho_P\le\rho_{em}\le\rho_{S}\le\rho_{MAX}$. We specialize the 
EMT to the case in which the random resistor network undergoes 
a metal-superconductor transition, by varying some control parameter, e.g., the 
temperature $T$, on which the resistivities $\rho_i$ depend. Lowering the 
temperature on the metallic side, an increasing number of resistors become superconducting and $\rho_{em}$ decreases. Near the transition  
$\rho_{em}\to 0^+$, and Eq.\ (\ref{effmed}) can be linearized as\cite{notaem} 
\[
\left[\frac{1+\alpha}{\alpha}\sum_{\rho_i>0}\frac{w_i}{\rho_i}\right]\rho_{em}
=1-(1+\alpha)\sum_{\rho_i=0}w_i,
\] 
whence it is evident that the transition occurs when the total weight of the 
superconducting resistors, $w_s\equiv\sum_{\rho_i=0}w_i$, equals $1/(1+\alpha)$. 
On a square lattice $\alpha=1$ and we find that the EMT correctly reproduces the 
percolation threshold $w_s=1/2$ in two dimensions.   

To gain further insight into the physics of the metal-superconductor transition
within the EMT, let us initially assume that the resistivity on a bond of the 
random resistor network may take only two constant values, $0$ and $\rho_0>0$,  
and that the network is characterized by a distribution of $T_c$'s 
$\mathcal W(T_c)$ of the individual resistors, with 
$\int_{-\infty}^{+\infty}dT_c \,\mathcal W(T_c)=1$. Henceforth, to simplify 
the notation, we adopt a continuous description of the statistical distributions.  
Each resistor has a resistivity $\rho_0$ at high temperature and its resistivity 
vanishes as soon as the temperature $T$ is lowered below its critical temperature 
$T_c$, i.e., the resistivity of a bond is taken as $\rho=\rho_0\theta(T-T_c)$, 
where $\rho_0$ is the same for all bonds and $T_c$ is a random variable. The 
distribution of resistivity in the system is therefore 
$w(\rho)=w_s(T)\delta(\rho)+[1-w_s(T)]\delta(\rho-\rho_0)$, where 
$w_s(T)\equiv\int_T^{+\infty}dT_c \,\mathcal W(T_c)$ is the statistical weight  
of the superconducting resistors at a temperature $T$, i.e., the frequency of 
occurrence of resistors with $T_c >T$. In such a case, Eq.\ (\ref{effmed}) may be 
readily solved, to yield\cite{kirkpatrick} 
\begin{equation}
\rho_{em}(T)=(1+\alpha)\rho_0\int_{T_c^\alpha}^{T}dT_c \,\mathcal W(T_c)
\label{solution}
\end{equation} 
for $T\ge T_c^\alpha$, and $\rho_{em}(T)=0$ for $T<T_c^\alpha$, 
where $T_c^\alpha$ is the critical temperature of the effective medium, 
which is defined by the equation 
\[
w_s(T_c^\alpha)\equiv\int_{T_c^\alpha}^{+\infty}dT_c \,\mathcal W(T_c)=\frac{1}{1+\alpha}.
\] 
Thus, for $T\to\infty$, $\rho_{em}\to\rho_0$, and $\rho_{em}(T_c^\alpha)=0$. 
On passing, we note the interesting inversion formula
$\mathcal W(T_c)=\rho_{em}^\prime(T)\vert_{T=T_c}/[(1+\alpha)\rho_0]$, which holds 
for  $T\ge T_c^\alpha$, and allows one to reconstruct the distribution of $T_c$'s 
for $T_c\ge T_c^\alpha$, from the behavior of the effective medium resistivity as 
a function of $T$. This relation holds provided that the normal-state resistivity 
$\rho_0$ does not depend on temperature. We apply the analytical solution 
(\ref{solution}) to a paradigmatic example which shall be used as a benchmark in 
the forthcoming analysis. Let us assume a distribution of $T_c$'s 
\[
\mathcal W(T_c)=\frac{w_1}{\sqrt{2\pi}\sigma_1}{\mathrm e}^{-(T_c-T_1)^2/2\sigma_1^2}
+\frac{w_2}{\sqrt{2\pi}\sigma_2}{\mathrm e}^{-(T_c-T_2)^2/2\sigma_2^2},
\] 
with $w_1+w_2=1$. This is a bimodal distribution of $T_c$'s, with two 
characteristic values, $T_1$ and $T_2$, and widths $\sigma_1$ and $\sigma_2$. As a  
limiting case, e.g., for $w_2=0$, a single Gaussian distribution is recovered. 
Specializing Eq.\ (\ref{solution}) to the present case, we find 
\begin{eqnarray}
\rho_{em}(T)&=&\frac{1+\alpha}{2}\rho_0\nonumber\\
&\times&
\left\{w_1\left[\mathrm{erf}\left(\frac{T-T_1}{\sqrt{2}\sigma_1}\right)
-\mathrm{erf}\left(\frac{T_c^\alpha-T_1}{\sqrt{2}\sigma_1}\right)\right]\right.\nonumber\\
&+&
\left.
w_2\left[\mathrm{erf}\left(\frac{T-T_2}{\sqrt{2}\sigma_2}\right)
-\mathrm{erf}\left(\frac{T_c^\alpha-T_2}{\sqrt{2}\sigma_2}\right)\right]
\right\},
\label{bimodal}
\end{eqnarray} 
where $\mathrm{erf}(x)=\frac{2}{\sqrt{\pi}}\int_0^{x}dz\,\mathrm e^{-z^2}$. 
The equation that fixes $T_c^\alpha$ is 
\[
w_1\,\mathrm{erf}\left(\frac{T_c^\alpha-T_1}{\sqrt{2}\sigma_1}\right)
+w_2\,\mathrm{erf}\left(\frac{T_c^\alpha-T_2}{\sqrt{2}\sigma_2}\right)=
\frac{\alpha-1}{\alpha+1}.
\] 
As it will be clear below, such bimodal distribution turns out to reproduce 
quite well a large tail in the resistivity as $T_c^\alpha$ is approached, 
as in Fig.\ \ref{figdeltaT}. Indeed, while for a single gaussian distribution 
both the EMT and the random-resistor network solution give a resistivity 
vanishing almost linearly at $T_c^\alpha$, when part of the system is not 
superconducting (i.e., $T_2$ vanishes) the transition occurs smoothly with a 
resistance vanishing with an upward curvature, in close resemblance with 
the experimental data in superconducting interfaces.\cite{triscone,espci}

\section{Exact solution of the random-resistor network}   
To have exact results and to test the reliability of the EMT, we solve numerically  
a system of resistors on a finite square lattice. Each link $(i,j)-(i+1,j)$ along 
$x$ or $(i,j)-(i,j+1)$ along $y$ of the lattice is characterized by a resistance 
which can vanish at a local critical superconducting temperature $T_c$. A 
given fixed voltage is applied at two opposite edges of the square $N\times N$ 
cluster, while the two other edges have open boundary conditions. Given a 
random distribution of resistivities and/or $T_c$'s on the links, the system 
adjusts the voltage at each node $(i,j)$ and the currents along each link 
by implementing Ohm's law on the links and Kirchoff's law for current conservation 
on the nodes. These laws provide a set of $3N^2-2N$ linear equations to be solved
to determine the $N^2-2N$ voltages of the nodes (the voltages on two sides of 
the cluster are fixed) and the currents of the $2N^2$ links (the inside $N^2-2N$ 
links plus the $2N$ incoming and outgoing links). Of course the system could be 
reduced to a smaller number of equations by simply inserting by hand the expressions 
of the voltages or of the currents obtained by Ohm's or Kirchoff's laws. However, 
in order to get a more transparent mapping of currents and potentials on each link 
or node, we choose this straightforward representation. We proceed as follows. 
We extract from a given distribution (which can be spatially uncorrelated 
or correlated) a critical temperature on each link. Given $T_c$ on a link, the 
normal-state resistivity is determined.  It can be taken as independent from $T_c$, 
or it can be larger for less metallic regions (i.e., those having a smaller $T_c$). 
The latter case is aimed to account for the effect of microscopic impurities on 
the critical temperature, and it will be discussed in Sec. IV-B within the context 
of the Finkelstein theory.\cite{finkelstein} The resistors on the links can even 
display a temperature dependence to describe, e.g., the contribution of  
superconducting fluctuations to the reduction of $R$ upon approaching the local 
$T_c$. This case will be discussed in Sec. IV-A within the context of both 
Gaussian (Aslamazov-Larkin) or vortical (Berezinski-Kosterlitz-Thouless) 
superconducting fluctuations. In any case, at a given temperature a given 
random realization of $T_c$'s gives rise to a realization of normal-state 
resistivities. Then the set of linear Ohm's and Kirchoff's equations is 
numerically solved for a $N\times N$ clusters with $N$ ranging from 50 to 200. 
The total current $I$ flowing at one edge of the cluster is evaluated by summing 
the currents of the $N$ horizontal links (if the overall constant voltage difference 
$V$ is applied to the two equipotential sides at $x=1$ and $x=N$). The ratio $V/I$ 
determines the total resistance of the cluster at any given temperature for 
the specific random realization of $T_c$'s and of the related local resistances in 
each bond $R(T_c,T)$. This exact numerical solution is then compared with the 
solution of the EMT.

\section{Spatially uncorrelated distributions of critical temperatures} 
\begin{figure}   
\includegraphics[angle=0,scale=0.3]{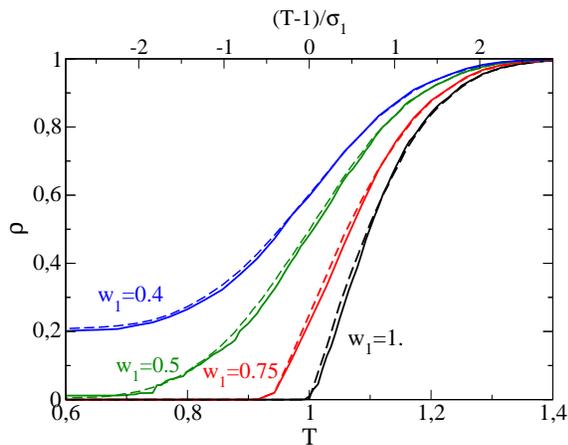}   
\caption{(Color online) Resistance curves for bimodal gaussian distribution of 
$T_c$'s. The solid lines are the exact results, the dashed ones report the 
EMT calculation. The upper scale is the deviation of $T$ from the average 
$\overline{T}_c$ in units of $\sigma_1$. The various colors refer to 
different weights of the two peaks of the bimodal distribution: black $w_1=1$; 
red $w_1=0.75$; green $w_1=0.5$; blue $w_1=0.4$. The upper peak of the bimodal
distribution has an average $\overline{T_c}=1$ and variance $\sigma=0.14$.}
\label{bimod}  
\end{figure}

We start considering the case of spatially uncorrelated distributions of critical 
temperatures. As we mentioned above, we will consider a bimodal distribution 
as a possible realization of smoothly vanishing resistivity curves. The 
$T_c$ distribution is then formed by two gaussian peaks: the first one with 
weight $w_1$ is centered at a positive $\overline{T}_c=T_1$ representing 
metallic regions with finite superconducting temperatures, while the second 
Gaussian with a weight $w_2$ represents regions with non-superconducting character 
and it is located at negative $T_c$'s (the precise location of this second Gaussian 
is immaterial as long as its tail is negligible on the positive $T_c$ side). 
Typically we chose ${T}_1=1$ and $\sigma_1\approx 0.14-0.16$. The relative 
weights $w_1$ and $w_2=1-w_1$ of the two components of the bimodal distribution 
tunes the ratio between regions which can and cannot become superconducting. Of 
course the case of a simple Gaussian distribution of superconducting temperatures 
is recovered by choosing $w_1=1$. Fig.\ \ref{bimod} reports the case where all 
the regions (both superconducting and non-superconducting) have the same 
normal-state resistivity $\rho_0=1$. As soon as $T$ decreases below the local 
$T_c$, the local resistor becomes superconducting and the local resistance is 
switched off. Starting from the purely Gaussian case $w_1=1$ (black solid
lines) we progressively reduce the weight of the metallic-superconducting regions 
to $w_1=0.75$ (online red lines), $w_1=0.5$ (online green lines), and 
$w_1=0.4$ (online blue lines). This has two effects on the $\rho(T)$ curves: (i) 
the critical temperature decreases, and (ii) the superconducting transition width 
increases, and the $\rho(T)$ displays a finite tail which eventually saturates to 
a finite value when $w_1< 0.5$. By defining the transition width as done in 
Fig.\ \ref{figdeltaT} one obtains the values reported in Table II. When $w_1=1$ 
the transition width can be estimated from Eq.\ (\ref{solution}), that gives the 
slope at the transition 
$\rho_{em}'(T_c)=2\rho_0 {\mathcal W}(T_c)=\sqrt{2/\pi}\rho_0/\sigma_1$, so that 
$\Delta T_c/T_c=\sqrt{\pi/2}(\sigma_1/T_1)$. Since we are using a distribution 
having $\sigma_1/T_1=0.14$ we obtain $\Delta T_c/T_c\approx 0.18$, as reported 
in Table II. However, when $w_1$ decreases $\Delta T/T_c$ increases, and 
simultaneously $\rho(T)$ displays a positive upward curvature after the regime 
of linear decrease.    
\begin{table}[t] 
\begin{center} 
\begin{tabular}{cc} 
\hline \hline 
$w_1$ & $\Delta T/T_c$ \\ 
\hline 
1 & 0.18 \\ 
0.75 & 0.28\\ 
0.5 & 0.5 \\ 
0.4 & 0.71 \\ 
\hline \hline 
\end{tabular} 
\end{center} 
\caption{Transition width for the bimodal distribution, according to
the results shown in Fig.\ \ref{bimod}.}
\label{p-table}
\end{table}   
As far as the comparison between the EMT (the dashed lines) and the numerical 
exact results is concerned, as it is apparent in Fig.\ \ref{bimod}, they coincide 
at any value of $w_1$, except in a very small critical region around the 
transition. This is quite remarkable since it is commonly believed that EMT works 
well only for dilute systems or for binary alloys of chemical species with 
similar resistivity.\cite{effectivemedium} Here the situation involves, instead, 
a binary alloy with finite ($=\rho_0$) and vanishing resistances at any 
relative concentration. 
We next explore the performance of EMT relaxing the assumption of constant 
normal-state $\rho_0$ for two cases of physical interest. In the first case a 
specific temperature dependence of the local resistances arises from 
paraconductive superconducting fluctuations inside each grain. In the second 
case quenched disorder induces a relation between the normal-state 
resistivities and the local critical temperature.    

\subsection{Resistance with paraconductivity fluctuations}    
\subsubsection{Gaussian superconducting fluctuations} 
\begin{figure}
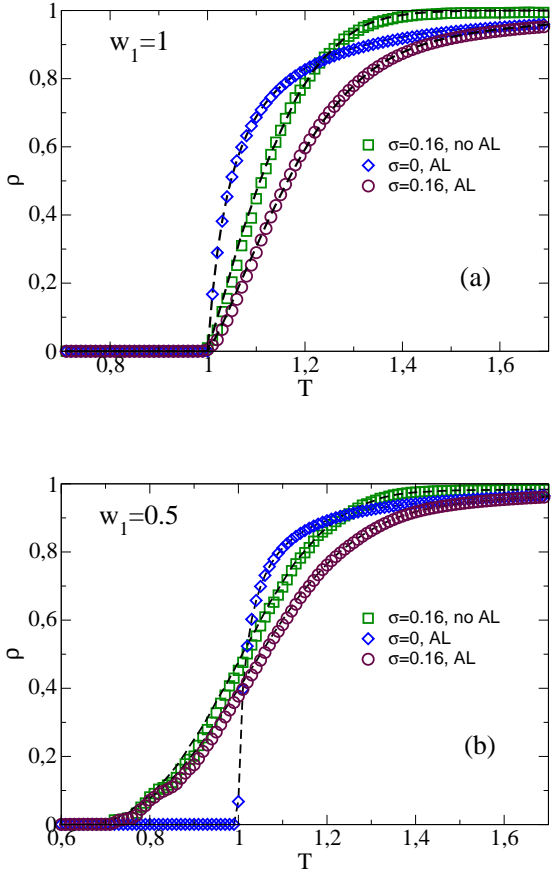
  
\includegraphics[angle=0,scale=0.3]{caprara-FIG3a.eps}  
\vskip 1.truecm
\includegraphics[angle=0,scale=0.3]{caprara-FIG3b.eps}  
\caption{(Color online) Resistance curves in the presence of AL paraconductivity 
fluctuations and/or disorder. Symbols represent the exact results, the dashed 
lines report the EMT calculation. (a) Gaussian distribution of disorder. 
(Squares) Gaussian distribution of $T_c$'s with $\sigma=0.16$ and $\overline{T}_c=1$
and no AL fluctuations ($a=0$); (diamonds) resistance in the presence of AL
fluctuations ($a=0.05$), but no disorder ($\sigma=0$); (circles) resistance in the 
presence of both AL fluctuations ($a=0.05$) and disorder ($\sigma=0.16$
and $\overline{T}_c=1$). (b) Same as in (a), but with a bimodal gaussian 
distribution of $T_c$'s with $w_1=0.5$.}
\label{ALfig}  
\end{figure}

In the case of gaussian superconducting fluctuations the well-known theory by 
Aslamazov and Larkin\cite{AL} introduces a power-law correction to the 
normal-state conductivity arising from Cooper-pair fluctuations. This increase 
of the conductivity diverges at $T_c$ with a prefactor which is universal in 
two dimensions even in the presence of disorder. Although other corrections 
to the conductivity become influential in the case of $s$-wave superconducting 
fluctuations, namely the Maky-Thompson contribution and the density of state 
corrections,\cite{varlamovlarkin} in two dimensions the overall importance of 
the paraconductive corrections is well estimated by the universal 
Aslamazov-Larkin coefficient. This gives rather precise indications on the 
strength of the paraconductivity corrections and allows to avoid unrealistic 
assumptions on the role of paraconductive fluctuations in shaping the 
resistivity curves. Fig.\ \ref{ALfig} reports two cases in which the 
normal-state resistivity is lowered by paraconductivity effects according to 
the Aslamazov-Larkin theory of superconducting fluctuations. In this case 
\begin{equation}
\label{asllark}
\rho_0(T)=\left(1+\frac{ aT_c}{T-T_c}  \right)^{-1}
\end{equation} 
when the local $T_c$ is lower than $T$, while it is zero as soon as $T$ becomes 
lower than $T_c$. Here $\rho_0(T)$ is the ratio of the resistivity to its 
normal-state value (so that $\rho_0(T)=1$ at high $T$), and within the ordinary 
AL theory the parameter $a$ is given in terms of the normal-state sheet resistance 
by $a=R_N/R_c$. Having in mind the experiments in the superconducting interfaces, 
where normal-state resistivity ranges bewteen $0.5$ and 
$2$ k$\Omega$,\cite{triscone,espci,sacepe1,sacepe2,sacepe3,mondal} we use as an 
indicative value $a=0.05$. In Fig.\ \ref{ALfig}(a) the critical temperatures 
have a Gaussian distribution, while Fig.\ \ref{ALfig}(b) reports the case of a 
bimodal distribution with weight $w_1=0.5$ of the gaussian peak with finite 
average $T_c$. Again the EMT (dashed lines) reproduces well the exact numerical 
data (symbols) in both cases. We notice in passing that, although the 
paraconductivity correction diverges in each grain (i.e., in each resistor 
approaching its $T_c$), the effect on the whole system is vanishingly small around 
the global $T_c$. This can be understood thinking that around the global $T_c$ 
only a minor fraction of the resistors is still normal: for the whole system it 
makes little difference whether the very last few resistors of the percolating 
cluster have their resistance constant or lowered by AL fluctuations. 
In general we notice that the paraconductivity corrections do lower the slope of 
the resistivity curves thereby increasing the width of the global superconducting 
transition (compare squares and circles). However, it is also clear that assuming 
a realistic paraconductivity strength one cannot account for the experimentally 
observed transition widths without considering also a substantial effect of 
disorder (see diamonds). Resistivity slopes as small as the observed ones (see 
Table I) can only occur if the width is ruled by disorder (i.e., by the width of 
the $T_c$ distribution), while AL paraconductivity can at most provide a 
moderate increase of the width.   

\subsubsection{Vortical superconducting fluctuations} 
\begin{figure}
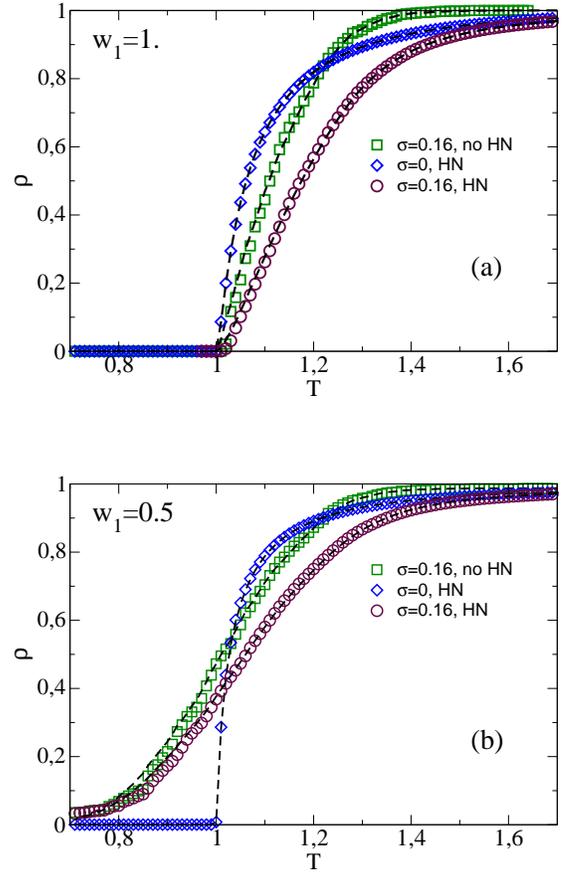
  
\includegraphics[angle=0,scale=0.3]{caprara-FIG4a.eps}  
\vskip 1.truecm
\includegraphics[angle=0,scale=0.3]{caprara-FIG4b.eps}  
\caption{(Color online) Resistance curves in the presence of vortical
paraconductivity fluctuations and/or disorder. The symbols  represent the
exact results, the dashed lines report the EMT calculation. (a) Gaussian 
distribution of disorder. (Squares) Gaussian distribution of $T_c$'s with 
$\sigma=0.16$ and no vortical fluctuations ($A=0$); (diamonds) resistance 
in the presence of vortical fluctuations [$t_c=0.1$, $\alpha=0.2$
($b=0.126$), and $A=2$], but no disorder ($\sigma=0$); (circles) resistance
in the presence both of vortical fluctuations [$t_c=0.1$, $\alpha=0.2$
($b=0.126$), and $A=2$] and of disorder ($\sigma=0.16$).
(b) Same as in (a), but with a bimodal gaussian distribution of $T_c$'s with 
$w_1=0.5$.}
\label{HNfig}  
\end{figure}

Owing to the two-dimensional character of the superconducting films, we also 
consider the possibility that paraconductivity corrections might arise from 
vortical fluctuations.\cite{HN} While the functional form of these corrections 
in terms of the coherence length $\xi$ is the same as for the Gaussian 
fluctuations, i.e., $\delta \sigma \propto \xi^2$, the prefactor is not universal 
and the temperature dependence reflects the exponential behavior of the 
coherence length in the Berezinski-Kosterlitz-Thouless (BKT) 
transition,\cite{HN,benfatto09} i.e., 
\begin{equation}
\rho_0(T)=\left[ 1+A\sinh\left(\frac{b}{\sqrt{t}}\right)\right]^{-1}
\label{rhn}
\end{equation} 
where $t\equiv (T-T_{BKT})/T_{BKT}$ is the distance from the BKT transition 
temperature $T_{BKT}$, $A$ is a constant of order one, and the parameter $b$ 
is controlled by the distance bewteen the $T_{BKT}$ and the mean-field 
temperature $T_c^0$, above which Gaussian fluctuations are restored. As it 
has been discussed in detail in Ref.\ \onlinecite{benfatto09}, $b=2\alpha
\sqrt{t_c}$, where $t_c\equiv (T_c^0-T_{BKT})/T_{BKT}$ and $\alpha$ is a measure 
of the energy of the vortex core, expressed in units of the value it assumes within 
the standard $XY$-model approach to the BKT transition. Having in mind these 
definitions, we first explored the effect of local vortical paraconductivity 
fluctuations on the global superconducting transition using $t_c=0.1$, 
$\alpha=0.2$ (i.e., $b=0.126$), and $A=2$, which correspond to realistic 
parameter values, according to the estimates for thin disordered superconducting
films in Ref.\ \onlinecite{benfatto10}. The results are shown in Fig.\ \ref{HNfig} 
where, similarly to what found above for the Gaussian paraconductivity effects, 
we find that the most realistic estimate of the parameters produces only rather 
small corrections to the widths of the transition.
It is quite apparent from Figs.\ \ref{ALfig} and \ref{HNfig} that, for a 
realistic choice of parameters, the effects of AL fluctuations and vortical 
fluctuations are quite similar in the absence of disorder as well as in the 
presence of both a gaussian or a bimodal distribution of $T_c$'s.

\begin{figure}  
\includegraphics[angle=0,scale=0.3]{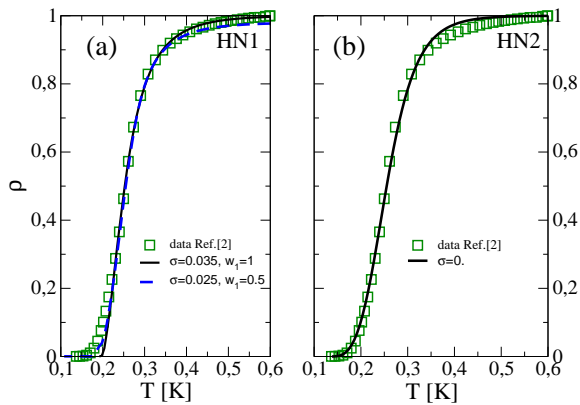}  
\caption{(Color online) (squares) Resistivity data of a LaAlO$_3$/SrTiO$_3$ 
sample at zero bias from Ref.\ \onlinecite{triscone}. (a) (Black) Solid line, fit 
in the presence of both gaussian disorder ($w_1=1,\, \sigma_1=0.035$ K, and 
$\overline{T}_c=0.195$ K) and vortical fluctuations (HN1) with $t_c=0.1|$, 
$\alpha=0.4$ ($b=0.25$), and $A=2$. The (blue online) dashed line is for a 
similar parameter set (HN1'), with $t_c=0.1$, $\alpha=0.4$ ($b=0.25$), and $A=1.8$, 
and a bimodal distribution $w_1=0.5$, $\sigma_1=0.025$ K, and $\overline{T}_c=0.23$ K.
(b) Squares, same as in (a); (black) solid line, fit in the presence of vortical
fluctuations only (HN2) with $t_c=0.7$, $\alpha=0.8$ ($b=1.34$), $A=0.71$, and
$T_c=0.157$ K.}
\label{HNfit}  
\end{figure}

Since in the literature the correct estimate of the parameters within 
the Halperin-Nelson formula (\ref{rhn}) is often debated,\cite{benfatto09} 
we decided to explore further the limit of the BKT theory in reproducing 
correctly experimental data in superconducting interfaces. In particular, we 
considered also a set of parameters able to yield a larger transition width, 
regardless of their microscopic determination. In Fig.\ \ref{HNfit} we show the 
results for two set of Halperin-Nelson parameters and the related fit to data 
from an unbiased sample in Ref.\ \onlinecite{triscone}. In panel (a) we 
report the results for a set (HN1) with $t_c=0.1$, $\alpha=0.4$ ($b=0.25$), and $A=2$,
in the presence of a gaussian distribution of $T_c$'s ($w_1=1$) with 
$\sigma_1=0.035$ K and $\overline{T}_c=0.195$ K (black solid line) and for a 
similar set (HN1') with $t_c=0.1$, $\alpha=0.4$ ($b=0.25$), and $A=1.8$, in 
the presence of a bimodal distribution of $T_c$'s ($w_1=0.5$) with 
$\sigma_1=0.025$ K and $\overline{T}_c=0.23$ K (blue dashed line). In panel 
(b), instead, we attempt to fit the same data with vortical fluctuations
only (i.e., $\sigma=0$). Of course we need a quite more substantial amount
of vortical fluctuations as given by a different (rather unrealistic\cite{notaNbN}) 
parameter set (HN2), with $t_c=0.7$, $\alpha=0.8$ ($b=1.34$), and $A=0.71$. 
It is apparent that vortical fluctuations alone can account for broad
transitions [and also produce the tail characteristic of the
LaAlO$_3$/SrTiO$_3$ (LAO/STO) or LaTiO$_3$/SrTiO$_3$ (LTO/STO) interfaces]. 
However, the parameters needed to produce the fit (which, by the way is quite 
poor in reproducing the high-temperature downward curvature) can hardly be 
justified on microscopic grounds. On the other hand, a moderate amount of 
disorder (notice that $T_c\approx 0.2$ K is in physical units here
and therefore $\sigma_1 \approx 0.03$ K is comparable with the values
$\sigma_1/T_c \approx 0.15$ of the figures, having normalized $\overline{T}_c=1$)
allows quite reasonable fits using a far more standard set of Halperin-Nelson 
parameters [see Fig.\ \ref{HNfit}(a)]. We find also in this case that the overall
width of the transition can be captured by the gaussian distribution
of $T_c$'s (with a moderate contribution of superconducting fluctuations),
but to reproduce the low-temperature tail one needs a bimodal distribution.

\subsection{Resistance with quenched disorder (Finkelstein's theory)} 
\begin{figure}   
\includegraphics[angle=0,scale=0.3]{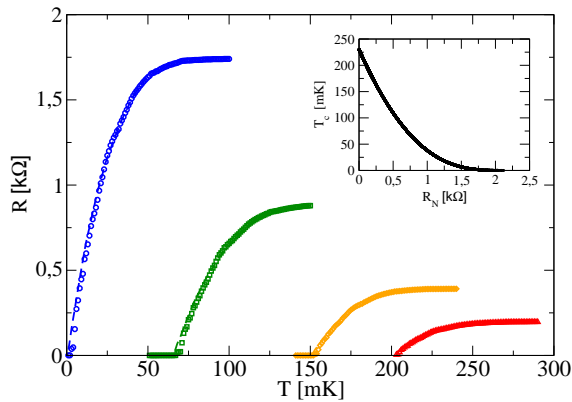}   
\vskip 1.truecm 
\caption{(Color online)  Resistance for systems with gaussian distribution of local 
$T_c$'s centered at different average $T_c$'s and related local resistances from 
Eq.\ (\ref{finkel}) with $R_c=81.1$ K$\Omega$, $T_c^0=230$ mK, and 
$\tau=5 \times 10^{-7}$ mK$^{-1}$ (the inset reports the $T_c$ vs. $\rho$ 
curve corresponding to Eq.\ \ref{finkel}). The variance of the gaussian distribution 
is always $\sigma=30$ mK. Symbols refer to exact calculations while dashed lines report the EMT curves. The gaussian distributions are centered
at $\overline{T}_c=0$ mK (blue circles and line), $\overline{T}_c=70$ mK
(green squares and line), $\overline{T}_c=150$ mK (orange diamonds and line),
and $\overline{T}_c=200$ mK, (red triangles and line).}
\label{gaussfink}  
\end{figure}

Another physically interesting case occurs when quenched impurities affect the 
critical temperature of the grains. In Ref.\ \onlinecite{finkelstein} a connection 
was established between disorder and the critical temperature in two-dimensional 
samples by deriving a relation between the resistance and the critical temperature 
in the presence of both disorder and Coulomb repulsion. Here, assuming that 
each grain is large enough to make the theory of Ref.\ \onlinecite{finkelstein} 
applicable, we extract randomly (from a Gaussian distribution) the local critical 
temperatures of the resistors and we consequently assign the local resistivity. 
In practice we numerically invert the relation 
\begin{equation}
\frac{T_c}{T_c^0}=e^{-\frac{1}{\gamma}}\left(
\frac{\gamma-t/4+\sqrt{t/2}}{\gamma-t/4-\sqrt{t/2}}
\right)^{\frac{1}{\sqrt{2t}}}
\label{finkel}
\end{equation} 
where, following the notation of Ref.\ \onlinecite{finkelstein}, $T_c^0$ is 
the critical temperature of the clean system, $t\equiv R_N / R_c'$, 
$\gamma\equiv 1/ln(T_c^0\tau)$, $\tau$ is the elastic scattering time, and
the resistance $R_c'\equiv 2 \pi^2 \hbar/e^2 =81.1$ K$\Omega$.
For the sake of concreteness, in our calculation we use $\gamma\approx -0.11$,
obtained choosing $T_c^0=230$ mK as the superconducting temperature for the 
clean system and $\tau =5 \times 10^{-7}$ mK$^{-1}$ for the value of the 
elastic scattering time. These values have been chosen to reproduce the correct 
orders of magnitude of experiments in LTO/STO samples.\cite{jerome} The 
resulting $T_c$ vs. The $R_N$ curve is reported in the inset of Fig.\ \ref{gaussfink}.
The results reported in Fig.\ \ref{gaussfink} again display a quite good agreement 
between the exact numerical calculations and the EMT calculations (the barely 
visible dashed lines).
We notice here that the progressive increase of the resistivity upon considering
more disordered systems with lower $T_c$'s is associated with an increasing slope 
of the resistivity curves near $T_c$. By rescaling all the curves to let them 
assume the same high-temperature value $\rho_0(T\gg T_c)=1$, the various curves 
are found to acquire quite similar slopes. This intriguing feature of our data 
is present in the case of LTO/STO interfaces, where, however, tails at 
$T\simeq T_c$ are also found, which are missing in our calculations of 
Fig.\ \ref{gaussfink}, carried out with a single gaussian distribution of local 
$T_c$'s.

\section{Spatially correlated distributions of critical temperatures} 
In all above cases the distribution of $T_c$'s and the consequent distribution of 
the resistances were spatially uncorrelated. This generically allows for a 
remarkably good performance of the EMT. Here, instead, we challenge EMT in the case 
of various space correlations.   

\subsection{Short-range averaged critical temperatures} 
\begin{figure}   
\vskip 1.truecm 
\includegraphics[angle=0,scale=0.3]{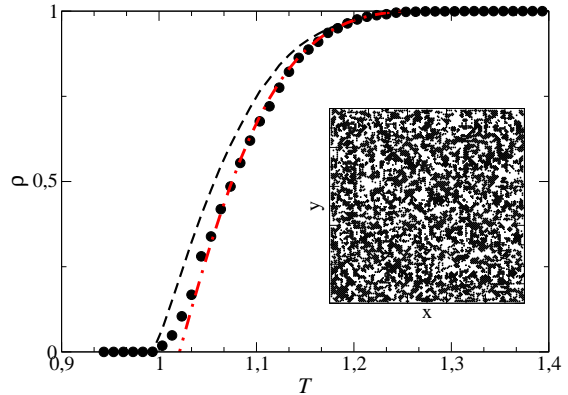}   
\caption{(Color online) Resistance of a short-range correlated random resistor 
network with $\overline{T}_c=1$ and a starting $\sigma=0.33$. The solids dots are 
the exact results, while the dashed line is obtained from numerical evaluation of 
the EMT, Eq.\ \ref{effmed}, with $\alpha=1$. The red dot-dashed line is instead 
obtained from Eq.\ \ref{bimodal}, with $w_1=1$, $w_2=0$, $\alpha=1.6$, $T_1=0.997$, 
and $\sigma_1=0.09$.} 
\label{correlgauss}   
\end{figure}

In order to investigate the effects of short-range space correlations, we implement 
a short-range averaging procedure of the critical temperatures. As a first step, we 
extract the local $T_c$'s of the bonds on our cluster from a given gaussian 
distribution. Then, the value of $T_c $ on each bond is averaged with the values on 
the six nearest neighbor bonds. Of course, this yields gaussian distributed $T_c$'s 
with a variance reduced by a factor $\sqrt{7}$ and, at the same time, creates space 
correlations on distances of the order of two lattice spacings. This short-range 
correlation can be extended by iterating the averaging procedure: the 
once-averaged $T_c$'s can be averaged again with the six nearest neighbors, leading 
to space correlations up to four lattice spacings. The results we display below 
are always obtained with a two-step averaging protocol. Calculations with one- 
or three-step averaging procedures yield similar results. Fig.\ \ref{correlgauss} 
reports with the solid black dots a calculation obtained starting from a 
Gaussian distribution with variance $\sigma=0.33$, centered at $\overline{T_c}=1$. 
The corresponding result for a EMT calculation is given by the black dashed line. 
Since EMT completely ignores the space correlations, the results are the same as 
those obtained with the starting uncorrelated set of random $T_c$'s, but for a 
trivial rescaling of the variance. Here, we numerically find that the variance 
entering the EMT expression [cf. Eq.\ (\ref{bimodal})] with $w_2=0$ is 
$\sigma_1=0.09$. Clearly the averaging procedure spatially correlates the 
high-$T_c$ regions, which form short-range clusters. The inset in 
Fig.\ \ref{correlgauss} displays a $100\times 100$ cluster in which the black 
regions correspond to regions where the local $T_c$ is larger than the average value 
$\overline{T_c}=1$. This lowers the exact resistance curve with respect to the  
EMT one, since the superconducting cluster is formed by objects with an 
effectively higher connectivity. Of course, this is true only far from percolation, 
because on a large scale it is instead immaterial whether the percolating objects 
are the single bonds or these small short-range clusters. Moreover, it is clear 
that the averaging procedure does not break the symmetry between high-$T_c$ and 
low-$T_c$ bonds (even if the variance is made smaller). Therefore, the percolation 
threshold stays the same and it is reached in two dimension when half of the bonds 
are superconducting. As a consequence, the curve of the exact calculation follows 
at high temperature the behavior of the EMT with a higher connectivity [this 
is represented by the (red online) dot-dashed line obtained from 
Eq.\ (\ref{bimodal}), with an effective connectivity numerically estimated to be 
$\alpha\approx 1.6$]. On the other hand, by lowering the temperature, the resistance 
is dominated by percolation effects, and a change of curvature occurs in 
$\rho(T)$, which vanishes at $T=1$, corresponding to a concentration 1/2 of 
superconducting bonds. This effect clearly introduces a ``tailish'' character
in the low-$T$ part of the resistivity, which, however, is not large enough 
to reproduce quantitatively the data in LAO/STO and LTO/STO interfaces.
\subsection{Patches} 
Besides averaging the $T_c$'s over the first neighbors, we also introduce space 
correlations by simulating the occurrence of patches, where the $T_c$'s are similar 
in a region of given size. We first select a set of sites randomly distributed in 
our $N\times N$ cluster. Around each of these ``seed'' sites we define a patch of 
a given radius $r$. Then, we extract the critical temperatures inside the patches 
from a given gaussian distribution centered at a higher value of 
$\overline{T_c}$, while the bonds outside the patches have vanishing ${T}_c$. In 
this way, we aim to simulate a bimodal distribution with spatial correlation over 
a range $r$ typical of a sample where different extended regions have a more or 
less marked superconducting character.  
\begin{figure}   
\includegraphics[angle=0,scale=0.3]{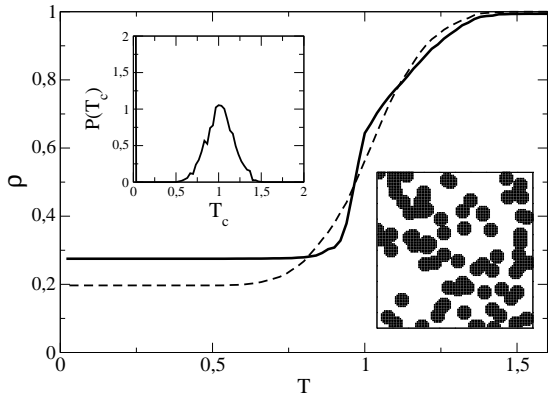}   
\caption{Resistance curves for a $100\times 100$ system with patchy structure 
(see lower inset). Each of the 80 patches (of radius $r=5$ lattice units) has a 
uniform local $T_c$ extracted from a Gaussian distribution with 
$\overline{T}_c=1$ and $\sigma=0.16$. The underlying matrix has local $T_c$'s 
extracted from a similar Gaussian centered at $\overline{T}_c=-100.$ (negative 
$T_c$'s are cumulated in $T_c=0$). The solid line is the exact result, while 
the dashed line is the EMT calculation. Upper inset: Total distribution of 
$T_c$'s.} 
\label{ni80}   
\end{figure}   
The lower inset of Fig.\ \ref{ni80} represents the case in which 80 patches 
with $r=5$ are introduced in a $100 \times 100$ cluster. The main panel of 
Fig.\ \ref{ni80} reports the related resistances both from exact (solid line) 
and EMT (dashed line) calculations. In both cases the system displays an initial 
rapid decrease of $\rho$, which then saturates at finite values upon decreasing 
$T$. While the EMT and exact data are quite similar at high temperatures, the 
low-$T$ saturation values are different. In particular the prediction of 
EMT is smaller than the exact result. Indeed, the exact solution is more 
resistive because the superconducting bonds are grouped into patches thereby 
leaving more extended surrounding regions, which are resistive. As a limiting case, 
one could imagine to form a single patch with weight 1/2 located inside the 
square cluster. Since one would have 1/2 of the bonds superconducting, 
EMT would predict the overall system to become superconducting. On the other hand, 
this single big patch being completely surrounded by a resistive region, the 
exact calculation would give a finite resistance for the system. This indicates 
that this realization of patches naturally results in higher resistances in the 
exact calculation, where the space correlation effects play a role, with respect 
to the EMT, where only the overall weight of superconducting bonds matters.  
In the present model, the size of superconducting regions can be increased either 
by increasing the number of patch centers (the ``seeds'') and/or by increasing 
the radius of the patches $r$. Fig.\ \ref{ni180} reproduces a case with 180 
patches with $r=5$ in a $100 \times 100$ cluster. Clearly (see lower inset), there 
is a large majority of highly superconducting sites, and the EMT predicts that  
the whole system becomes superconducting at $T_c=0.91$ (black dashed line in the 
main panel). Remarkably, this prediction reproduces well the exact solution given 
by the black solid line. However, close inspection of the lower inset shows that 
the superconducting character of the exact solution is only due to a small 
overlap of the patches in the central region of the cluster. If this small 
superconducting region were absent, the whole assembly of patches would not 
percolate and the system would display the finite resistance due to this small 
resistive region.
\begin{figure}   
\includegraphics[angle=0,scale=0.3]{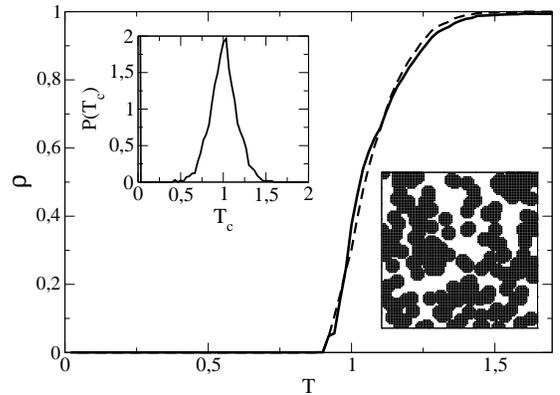}   
\caption{Same as in Fig.\ \ref{ni80}, but for a 
system with $180$ patches.}
\label{ni180}   
\end{figure}   
The above patchy structures have a random character, which makes it difficult 
to unambiguously establish the effective connectivity of the correlated regions. 
Therefore, we also investigated a toy model where the patches no longer form 
around random centers, but they form in a rather regular way. Specifically,  
we simply subdivided our $N\times N$ cluster into smaller $(N/d)\times (N/d)$ 
squares. This gives rise to a checkerboard support with $d\times d$ squares each 
having a given $T_c$ randomly extracted from a gaussian distribution. As a 
consequence, the critical temperatures of this coarse-grained system are still 
gaussian distributed but also display a strong space correlation, since 
they all coincide within each of the $d\times d$ square subclusters. 
We also notice in passing that to obtain a reasonably good Gaussian 
distribution quite large values of $N$ are required, because the statistical 
distribution of $T_c$'s is substantially degraded by extracting just 
one random $T_c$ inside each subcluster. We notice, however, that 
in this toy model each square subcluster is only connected to four 
other subclusters (but for a small effect in the corners). The exact 
calculation (not shown here) in this case agrees remarkably well with the 
prediction of the EMT. This demonstrates that connectivity plays a much more 
important role than the simple space correlation. Of course, whenever 
this latter influences the connectivity (like in the models of Secs. 
V A and V B), space correlation again becomes a main actor of the game.    

\subsection{Long-range spatially correlated distributions} 
\begin{figure}  
\includegraphics[angle=0,scale=0.7]{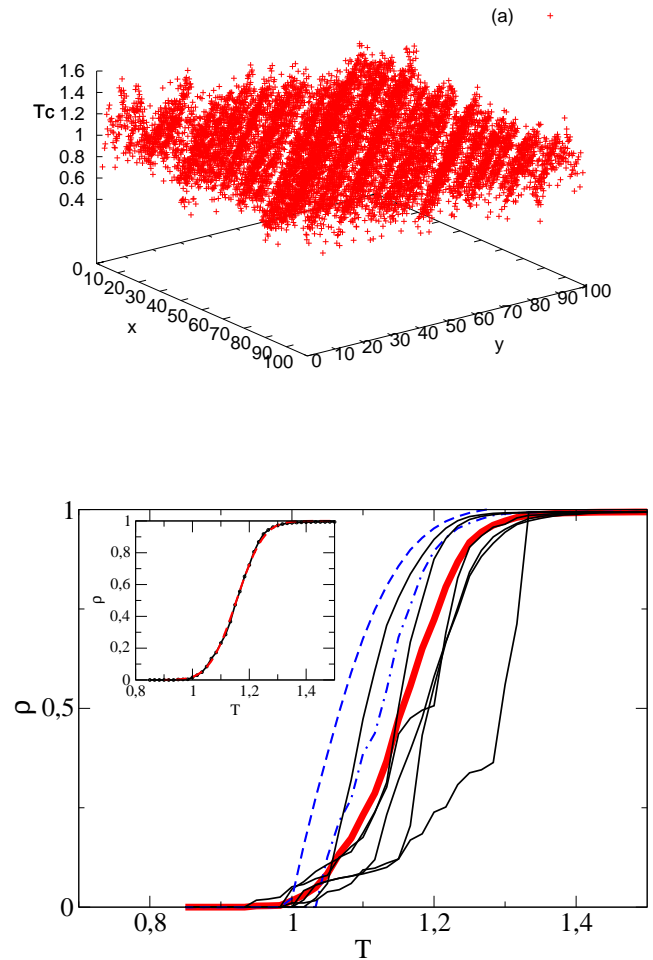}  
\vskip 1.truecm
\includegraphics[angle=0,scale=0.35]{caprara-FIG10b.eps}  
\caption{(Color online) (a) Example of a highly spatially correlated realization
of random Gaussian distribution of $T_c$'s with $\overline{T}_c=1$ and 
$\sigma=0.1$. (b) Resistance curves for various realization of similar 
random spatially correlated $T_c$'s (thin solid black lines) and their average
(thick solid red line). One of these specific curves has been drawn with a blue 
dot-dashed line together with its numerically calculated EMT (dashed line). 
Inset: The average resistance and a fit from Eq.\ (\ref{bimodal}) with $w_1=1$,
$w_2=0$, $\alpha=0$ (corresponding to an effective dimensionality $D=1$), 
$T_1=1.155$, and $\sigma_1=0.075$.
}
\label{correlgaussLR}
\end{figure}

To complete the analysis on the reliability of EMT in the presence of spatial 
correlations, we also investigate the behavior of the exact resistance and EMT  
when long range correlations are present. We empirically simulate this situation 
by adjusting the seeds of the random-number generator producing the distribution 
of random $T_c$'s, while visiting the $N\times N$ lattice sites, so that the 
$T_c$'s are highly correlated along lines. In particular, one can easily introduce 
long-range correlated patterns like the one reported in 
Fig.\ \ref{correlgaussLR}(a), where high values of $T_c$ along diagonals alternate 
with low values forming a stripe-like texture. In this case, the exact resistance 
displays rapid drops whenever the temperature reaches a value at which several 
neighboring bonds become simultaneously superconducting, substantially extending 
the size of the previously formed superconducting clusters 
[see the thin solid lines in Fig.\ \ref{correlgaussLR}(b)]. 
On the other hand, it then becomes difficult to build a fully percolating cluster 
and more or less sizable tails are generated. Lacking any information about 
the space structure of the random system, EMT fails in reproducing this situation. 
This is clearly visible comparing the two blue lines in Fig.\ \ref{correlgaussLR}(b),
reporting the exact (solid) and EMT (dashed) results for the same specific 
random realization of Fig.\ \ref{correlgaussLR}(a).

The sharp drops occurring in the individual disordered realizations can however 
be smoothed by averaging over various realizations of the long-range correlated 
patterns (simulating, for instance, a system with differently oriented textures).
The thick solid (red online) line in Fig.\ \ref{correlgaussLR}(b) is indeed 
obtained by averaging over the thin solid resistance curves obtained 
from seven different realizations with differently oriented and correlated
diagonal stripes like the one reported in Fig.\ \ref{correlgaussLR}(a).
As mentioned above, each of these realizations has a resistance quite different 
from that obtained with EMT. The (blue online) dashed line is the the EMT 
result obtained with the same gaussian distribution of local $T_c$'s used in 
Fig.\ \ref{correlgaussLR} ($w_1=1$, $w_2=0$, $T_1=1$, and $\sigma_1=0.1$). 
We also fitted the average curve [see inset in Fig.\ \ref{correlgaussLR}(b)]
by the analytic form in Eq.\ (\ref{bimodal}). This time the fit indicates an 
effective reduction of the connectivity, with $\alpha=0$ (i.e., $D=1$), 
$\overline{T}_c=T_1=1.15$, $\sigma_1=0.075$. While it is rather natural that 
some generic one-dimensional character arises from spatial correlations like in 
Fig.\ \ref{correlgaussLR}(a), the fact that the averaging procedure transforms 
the various quite different resistances of the individual realizations into a 
single nearly one-dimensional resistance curve is interesting.

\section{Conclusions}     
In this paper we aimed to understand whether mesoscopic inhomogeneities 
could provide the physical mechanisms producing unusually broad superconducting 
transitions in various nearly two-dimensional systems, like the 
quasi-two-dimensional electron gases formed at the interfaces between SrTiO$_3$ 
and LaAlO$_3$ or LsTiO$_3$. Furthermore we wanted to get insights on the 
``tails'' appearing in the resistance curves of LTO/STO and LAO/STO interfaces. 
By modeling the inhomogeneous system by a random resistor network, we 
systematically investigated various realizations of random resistors, which we 
solved both by an exact numerical approach and by a standard EMT approach. We 
started with the simple assumption of a random distribution of local $T_c$'s 
with global superconductivity occurring when superconducting regions (links) 
will eventually percolate by lowering the temperature. We then discussed 
various physical ingredients to account for the main experimental features of 
the superconducting transition at the interfaces. First of all, the transitions 
in these systems are anomalously broad and we clarified that this can hardly 
be attributed to standard gaussian or vortical (i.e., BKT-like) 
paraconductivity fluctuations. This is particularly evident for the 
AL paraconductivity, but it also holds for the Halperin-Nelson formula. Indeed, 
the set of parameters to be used to fit the experiments without disorder
comes out to be rather unphysical. Therefore, superconducting fluctuation 
effects can contribute to, but cannot completely account for, the substantial 
width of the transition and for the small slopes of $R(T)$ around 
$T_c$.  On the contrary, within our model the width of the transition 
mainly stems from the width of the distribution of $T_c$'s among the various 
superconducting mesoscopic regions. It turns out that a typical gaussian 
distribution of local $T_c$'s has the right downturn to mimic the experimental 
resistance curves, which display a moderate rounding above $T_c$ followed by a 
rather broad, nearly linear decrease when $T$ is lowered.  

We also found in passing that the degree of reliability of EMT in solving 
different disorder realizations is high and standard EMT generically reproduces 
quite well the exact resistance curves, whenever the space correlations between 
the mesoscopic regions are negligible. This result is general, and is not limited 
to dilute systems or to mixtures of similar resistances. We rather find that 
EMT works well for mixtures of metallic (finite $R$) and superconducting (i.e., 
$R=0$) resistances, for any filling ratio, possibly including the effects of 
superconducting fluctuations. On the contrary, we find that increasing the 
space correlations leads to more substantial failures of the EMT.    

An additional interesting and specific feature of the two-dimensional electron 
gas formed at the interface of the SrTiO$_3$ substrate is a more or less long tail 
in the low-$T$ part of the $R(T)$ resistance curves. These tails can even flatten 
to form a finite-$R$ plateau in cases where the system stays non superconducting  
despite a substantial decrease of $R(T)$ indicating the presence of a sizable 
superconducting fraction of the two-dimensional gas. We found that this feature is 
not easily reproduced and we only found few specific cases, where it could be 
reproduced within our model. In the case of non spatially correlated local $T_c$'s 
we found that a tail is only present when a bimodal distribution of $T_c$'s is 
assumed, with (or very near to) the specific $1/2-1/2$ distribution of the 
relative weight of the two distribution peaks. Since in the real systems the tails 
are observed over a broad range of fillings, magnetic fields, bias, it is hard 
to understand how these different conditions can all correspond to such 
specific $1/2-1/2$ constraint on the distribution of the mesoscopic local $T_c$'s. 
The $1/2-1/2$ condition can be rephrased by saying that at percolation, the whole 
set of superconducting domains randomly distributed in the two-dimensional system 
has to be exhausted.\cite{nota1D} One possibility which deserves to be explored 
is that some microscopic mechanism acts to create some low-dimensional skeleton 
on which superconductivity takes place. Localization effects have already been 
proposed as a possible mechanism to generate a fractal substrate for 
superconductivity\cite{feigelman} and a glassy superconducting 
phase.\cite{ioffe,feigelman10} If superconductivity occurred on this 
low-dimensional substrate, most if not all the bonds of the substrate have to 
be superconducting before the whole ``skeleton'' acts as a whole superconducting 
path.    

A further mechanism to explore in order to reproduce tails in the resistance of  
the quasi-two-dimensional electron gas is provided by space correlations (indeed, 
the above scenario of fractal or glassy superconducting phase is a form of space 
correlation). When space correlations are short-ranged (see Sec. V A), we detect 
the presence of a short tail (see Fig.\ \ref{correlgauss}). One can expect that 
extending the strength and the range of these correlation could result into 
a strengthening of the tail feature. This expectation is confirmed by investigating 
space structures where more metallic domains cluster forming patchy textures 
embedded into less metallic matrices with very low or zero superconducting 
temperatures (see Figs.\ \ref{ni80} and \ref{ni180}). In these cases again bimodal 
$T_c$ distributions are formed and again ``tailish'' resistances are obtained 
whenever the weight of the high-$T_c$ distribution peak approaches one half (or 
less when a plateau is obtained). 

Finally, to reproduce tailish resistance curves, we explored the case of textured 
domains (forming, e.g., stripe-like paths) with long-range space correlations. If 
the real system is formed by a patchwork of randomly oriented stripe domains, 
one would obtain resistance curves with substantial tails (see 
Fig.\ \ref{correlgaussLR}). Indications of a spatially structured system hosting 
the two-dimensional electron gas have indeed been obtained from scanning 
tunneling microscopy experiments on LAO/STO samples,\cite{salluzzo} suggesting 
that also this possibility is worth being further explored.   

\vskip 0.5truecm 
\par\noindent 
{\bf Acknowledgments.} 
We are indebted with N. Bergeal, C. Di Castro, J. Lesueur, and J. Lorenzana,   
for interesting discussions and useful comments. We acknowledge financial 
support from MIUR Cofin 2007, Prot. 2007FW3MJX\_003.


\begin{thebibliography}{99}
\bibitem{notaem} This is true for $\alpha>0$. For $\alpha=0$, in a one-dimensional 
network, Eq.\ (\ref{effmed}) has the exact analytical solution 
$\rho_{em}=\sum_i w_i\rho_i\equiv\rho_S$, whence it is evident that the network 
becomes superconducting only if all the resistors become superconducting.
\bibitem{notaNbN} For example, in Ref.\ \onlinecite{benfatto10} the largest 
$t_c$ value obtained in a three-nanometer thick film of NbN is of order
of $t_c=0.1$.
\bibitem{nota1D} A similar effect would occur in a one-dimensional system, where the 
superconducting transition takes place only when all the bonds (resistors) have 
become superconducting.
\end{thebibliography}
\end{document}